\documentstyle[prl,aps,epsfig,multicol]{revtex}

\begin{document}

\title{Possible spin triplet superconductivity in 
Na$_x$CoO$_{2}\cdot y$H$_{2}$0}

\author{ 
Akihiro Tanaka and Xiao Hu
}
\address{
Computational Materials Science Center, 
National Institute for Materials Science, Tsukuba 305-0047, Japan
}

\date{April 17, 2003}

\maketitle

\begin{abstract}

Combining symmetry based considerations with inputs from 
available experimental results, we make the case that 
a novel spin-triplet superconductivity triggered by antiferromagnetic fluctuations 
may be realized in the newly discovered layered cobaltide 
Na$_x$CoO$_{2}\cdot y$H$_2$O. In the proposed picture,  unaccessable via   
resonating-valence-bond physics extrapolated from half-filling,  the pairing process is similar to 
that advanced for Sr$_{2}$RuO$_4$,  but enjoys a further advantage coming from the 
hexagonal structure of the Fermi-surface which gives a stronger 
pairing tendency.

\end{abstract}

%\pacs{PACS number: 13.20.He, 11.30.Er, 12.15Hh, 11.80Et}

%\maketitle

\begin{multicols}{2}[]

%\noindent
Anderson's suggestion     
that a doped antiferromagnet on a 2d square lattice 
be best described in terms of resonating valence bond (RVB) states\cite{Anderson RVB}, 
set the direction for a huge subsequent effort pursuing this picture in an attempt to  
understand %the mechanism of high temperature 
superconductivity 
in the cuprate oxide compounds. While this has lead to 
the development of new avenues in the physics of low dimensional 
condensed matters, conclusive evidence of such novel states 
has not turned up. 
Meanwhile, renewed interests in RVB physics in systems on {\it triangular} 
lattices (TL) have been emerging recently - though the  
S=1/2 antiferromagnetic (AF) Heisenberg model on a TL  
is generally regarded as having a ground state with long range order, closely related systems 
%seem to 
%indicate a certain amount of stability of RVB-like states. Moessner and Sondhi 
%have shown
such as the undoped quantum dimer model\cite{Moessner} 
%that the undoped quantum dimer model posesses a finite range 
%of parameters (in contrast to the square lattice counterpart) in which 
%the short range RVB ground state is realized. Koretsune and Ogata \cite{Koretsune-Ogata}have found 
%strong indications that 
and the $t-J$ model on a TL with the hopping sign %has RVB feautures for the case 
$t > 0$\cite{Koretsune-Ogata} have been shown to exhibit a certain amount of stability of RVB-like states.

The discovery by Takada et al of superconductivity in 
Na${}_{x}$CoO$_{2}$ $\cdot y$ H$_{2}$O ($x\approx 0.35, y\approx 1.3$)\cite{Takada}
, a system made of stacked layers of cobalt ions 
sitting on a 2d TL and surrounded by 
octahedrally displaced oxygen atoms will no doubt spark new interest in this 
line of investigation. Superconductivity was achieved at T${}_{c}$=5K 
by oxidation of  the sodium ions (changing the effective charge 
density) followed by intercalation of the CoO$_2$ layers in the mother 
compound NaCo$_{2}$O$_{4}$ with H$_2$O, thus effectively 
enhancing the two dimensionality of the system. Guided by analogy with the 
cuprates, the authors of ref\cite{Takada} were lead to suggest that 
this system maybe understood in terms of doped AFs on TLs, 
which may well be related to RVB states.

As much as such directions are tempting, 
one must proceed with due caution since the experimental situation is rather complex. 
As discussed shortly, the system is actually away from the vicinity of half-filling, the regime where RVB physics is 
considered to be valid.  It is reasonable then to resort to generic methods  
no more specialized than a combination of     
symmetry analysis and simple ^^ ^^ fermiology",  reinforced by experimental inputs. 
While experiments conducted on  
Na${}_{x}$Co$_{2}$ $\cdot y$ H$_{2}$O are at present basically limited to resistivity and 
preliminary susceptibility data, we have noticed that several conclusions -which are in fact 
remote from what an RVB-based 
investigation would lead to- can already be drawn.  
We therefore believe it useful at this stage to raise notice on this point  
with the hope of motivating further activity.   
The seminal work of Rice and Sigrist, who adopted a similar approach 
in the case of Sr$_{2}$RuO$_{4}$\cite{Rice},  had played an instrumental 
role in the subsequent identification of the triplet superconductivity in 
that compound.  

During the final stage of this work, we came across a preprint by 
Baskaran\cite{Baskaran}
addressing the same system. 
Although his reduction of the problem to a one-band model 
is similar, the discussion there is based on the RVB picture. The present 
work is founded on a complimentary approach as explained above.

\noindent
{\it Deduction of effective theory} 

The first point to make is that the present system {\it cannot} be taken as a  
realization of the dilutely doped AF on a TL. 
The 3d cobalt ions in the parent compound NaCo$_{2}$O$_{4}$
with a formal valence of +3.5, are actually  
a mixture of the low-spin 
states Co${}^{3+}$(t${}_{\rm 2g}^{6}$e${}_{\rm g}^0$ with S=0) and 
Co${}^{4+}$(t${}_{\rm 2g}^{5}$e${}_{\rm g}^0$ with S=1/2). 
With no experimental indications of charge ordering 
in NaCo$_{2}$O$_{4}$, the most natural starting point for studying these 
systems is as a random mixture (1:1) of S=0 and S=1/2 states distributed 
on the TL. Hence instead of starting near half-filling, 
we must consider the case of three-quarter filling, which is a totally new 
situation. 
%t is this randomness and its associated enhancement of entropy 
%which has lead to the unusally large thermal power 
%in NaCo$_{2}$O$_{4}$\cite{Terasaki}. 
Superconductivity occurs after 
oxidation of the sodium ion so that the 
compound becomes Na${}_{0.35}$CoO$_{2}$ $\cdot y$ H$_{2}$O, in which case 
the system is approximately 1/3 filled. 
We therfore stress that RVB physics, presumably  
valid close to half-filling,  by no means has a trivial  
extrapolation to this regime.

The  next issue to consider is the magnetic properties of the system. 
The susceptibility\cite{Ono} of NaCo$_{2}$O$_{4}$  
contains a Curie-Weiss component with large 
negative Curie temperature T$_{\rm Curie}\approx -285K$, a 
trend also confirmed in measurements on 
Na${}_{0.35}$CoO$_{2}$ $\cdot y$ H$_{2}$O, though with a 
reduced value of T$_{\rm Curie}\approx -18.8K$\cite{Sakurai}. Taken together with the 
finding of spin density wave (SDW) like tendencies upon 
Co to Cu subsitution in NaCo$_{2}$O$_{4}$\cite{Terasaki SDW}, it is natural 
to assume a fair amount of AF fluctuation to be present. 
 LDA calculations show that the Fermi surface for NaCo$_{2}$O$_{4}$ 
 sits near the top of the a$_{1g}$ like state, split off from  
 the t$_{2g}$ manifold due to oxygen distortions\cite{Singh}. These circumstances suggest  a one-band Hubbard model ($U\approx 5-8eV, t\approx 1eV$) on a TL around quarter to third filling as the simplest model capturing the essential physics of this system. 
 At this point one may already expect nontrivial pairing tendencies. For instance, there are perturbative 
 studies that claim that triplet pairing is favored over singlet pairing\cite{Nisikawa}.   
With direct quantum Monte-Carlo methods unavailable due to 
 severe negative signs,  
 we proceed with the generic approach mentioned earlier.

\noindent{\it Symmetry considerations}

Group theoretical studies, valid irrespective of the pairing mechanism, 
have played essential roles
in classifying superconductivity in heavy fermions, organics, and e.g. the 
ruthenate compound\cite{Sigrist-Ueda,Mineev}.  
For the present case, the relevant symmetry group is   
${\cal G}=D_{6h}\otimes U(1) \otimes T$, with $D_{6h}$ the hexagonal group 
together with the inversion symmetry about the plane, $U(1)$ the gauge symmetry broken by the onset of superconductivity, and $T$ the time reversal symmetry. For Na${}_{0.35}$CoO$_{2}$ $\cdot y$ H$_{2}$O, 
magnetization measurements  
suggest the presence of a magnetic anisotropy\cite{Sakurai}.    
Hence the absence of $SU(2)$ from  
${\cal G}$. The irreducible representation for this situation can be classified 
into 12 classes, ${\Gamma}^{\pm}_{a}, 1\le a \le 6$, with +(-) standing for 
spin singlet (triplet) states. States with subscripts $1\le a \le 4$, and those with $a=5, 6$ are 1d and 2d 
representations respectively. The conventional point of view\cite{Nakajima} that AF fluctuations 
are most compatible with singlet pairing will lead us to one of these (+) representations. 

A more intriguing possibility however is the case of 
triplet pairing. Not easily accessed from RVB treatments which deal primarily with spin singlets 
and their fluctuations, 
it is perhaps worthwhile to highlight their exotic properties. Later on we will mention that 
experiments so far seem to spell out such a state as the one most probable, and  a novel scenario from which 
we recover this pairing choice will be described. 
As mentioned there,  we will be interested in  
states  
with the d-vector perpendicular to the cobalt plane, i.e. ${\vec d}//{\hat z}$, 
which ensures that the electron spins contributing to the wavefunctions 
have a vanishing z-component. 
It turns out that the 
$\Gamma^{-}_5$ representation realizes such states. 
The fourth order terms of a Ginzburg Landau expansion of the free energy 
is of the form 
\begin{equation} 
{\cal F}_{quad}=\int d^2 {\vec r}
[\beta_{1}(\vert\eta_{1}\vert^2 +\vert\eta_{2}\vert^2)^2 
+\beta_2 (\eta_{1}^{*}\eta_2 -\eta_{1}^{*}\eta_{2})^{2}],  
\end{equation} 
where $\eta_1$ and $\eta_2$ are coefficients of the two basis 
functions ${\hat z}k_x$ and ${\hat z}k_y$. 
When the coefficient $\beta_2 > 0$,  we have 
${\vec d}={\hat z}(k_{x}+ik_{y})$, a state violating $T$. 
The same pairing state 
also arises in the tetragonal ($D_{4h}$) case, though additional conditions on the GL coefficients 
are required there.  
The  $T$-violation is most dramatically 
manifested in electric/magnetic properties. 
The derivative terms of the GL theory 
for representations $\Gamma^{\pm}_{5}$ are 
\begin{eqnarray}
{\cal F}=\int d && {\vec r}  [K_{1}\vert D_{x}\eta_1 +D_y \eta_2 \vert^2 
+ K_2 \vert D_x \eta_2 - D_y \eta_1 \vert^2 \nonumber \\
&&+ 
K_{3}\left(
\vert D_{x}\eta_1 -D_y \eta_2 \vert^2 +
\vert D_{x}\eta_2 +D_y \eta_1 \vert^2
\right)]
%\nonumber \\
%&&+ 
%K_{4}(\vert D_{z}\eta_1 \vert^2 +\vert D_{z} \eta_2 \vert^2 )],
\end{eqnarray}
with $D_{\mu}=\partial_{\mu}-2ieA_{\mu}$ the covariant derivatives.  
All z-derivatives are omitted on account of the two-dimensionality. 
Adapting a method due to Furusaki et al\cite{Furusaki} who discussed the tetragonal case $D_{4h}$, we find that the terms in the GL Lagrangian density 
coupling linearly to the 
electric field ${\vec E}$ read
\begin{eqnarray}
{\cal L}_E &=&
%c_1(E_x \eta_{x}^{*}D_{x}\eta_{x}+ E_{y}\eta_{y}^{*}D_{y}\eta_{y})\nonumber\\
%&+&c_2(E_x \eta_{y}^{*}D_{x}\eta_{x}+ E_{y}\eta_{x}^{*}D_{y}\eta_{y})\nonumber\%\
%&+&c_3(E_x \eta_{x}^{*}D_{y}\eta_{y}+ E_{y}\eta_{y}^{*}D_{x}\eta_{y}
%)\nonumber\\
%&+&c_4(E_x \eta_{y}^{*}D_{y}\eta_{x}+ E_{z}\eta_{y}^{*}D_{y}\eta_{y})\nonumber\%\
%&+&c_5(E_x \eta_{x}^{*}D_{x}\eta_{x}+ E_{y}\eta_{y}^{*}D_{y}\eta_{y})\nonumber\%\
%
c_1 (E_x \eta_{1}^{*}+E_{y}\eta_{2}^{*})(D_{x}\eta_{1}+D_{y}\eta_{2})+c.c.
\nonumber\\
&+&
c_2 (E_x \eta_{2}^{*}-E_{y}\eta_{1}^{*})(D_{x}\eta_{2}+D_{y}\eta_{1})+c.c.
\nonumber\\
&+&c_{3}\{
 (E_{x} \eta_{1}^{*}-E_{y}\eta_{2}^{*})(D_{x}\eta_{1}-D_{y}\eta_{2})
\nonumber\\
&&\hspace*{5mm}+
(E_x \eta_{2}^{*}+E_{y}\eta_{1}^{*})(D_{x}\eta_{2}+D_{y}\eta_{1})+c.c.\}
%\nonumber\\
%&+&(E_{z} -terms),
\end{eqnarray}
where the coefficients $c_{j}$ (1$\le$j$\le$3) are proportional to 
the $K_j$'s in the free energy. 
From these one extracts the following contribution; 
\begin{eqnarray}
{\cal L}_{\rm CS-like}&=&i(c_{3}-c_{1}-c_{2})
(\eta_{1}^{*}\eta_{2}-\eta_{2}^{*}\eta_{1}) \nonumber\\ 
&& \times(A_{y}\partial_{x}A_{0}-A_{x}\partial_{y}A_{0}), 
\end{eqnarray}
which is of the Chern-Simons (CS) form. Several consequences follow, which are basically 
similar to what has been discussed in the 
context of SrRu$_2$O$_4$. 
First, a zero field Hall effect is expected, due to 
a spontaneous (orbital) magnetization 
${\vec d}\cdot({\vec k}\times\nabla_{\vec k}){\vec d}\propto {\vec \eta^{*}}\times{\vec \eta}$, where 
${\vec  \eta}\equiv (\eta_1 , \eta_2 )$. Futhermore one sees from 
the nonuniversality of the coefficient of the CS term that 
the Hall conductivity $\sigma_{xy}$ is not quantized. 
The physical reason for the 
latter feature can be identified microscopically. (Readers are referred to 
refs.\cite{Goryo,Senthil SQHE} for the tetragonal case.) 
Inserting the 
d-vector ${\vec d}=\Delta_{0}(f_1 ({\vec k}) +if_2 ({\vec k}) ){\hat z}$, 
with $f_1 ({\vec k})=2\sin(\frac{\sqrt{3}}{2} k_x) 
\cos(\frac{1}{2} k_y)$ and 
$f_2 ({\vec k})=2\cos(-\frac{\sqrt{3}}{2} k_{x})\sin({\frac{1}{2}k_y})
+2\sin k_y$ into 
the Bogoluibov-de Gennes Hamiltonian 
\begin{equation}
H={\Psi}^{\dagger}[
(\epsilon(\vec k)-\mu)\tau_{3}\otimes {\bf 1}+
\tau_1\otimes i({\vec d({\vec k})}\cdot{\vec \tau})\tau_2
]{\Psi},
\end{equation} where 
$\epsilon({\vec k})=-4\cos(\frac{\sqrt{3}}{2}k_x )\cos(\frac{1}{2}k_y )
-2\cos(k_y)$
and performing a gradient expansion to one-loop order, one sees that the 
correction to the universal contribution $\sigma_{xy}=e^2 /4\pi$ 
comes from the normalization of the generating functional 
due to the Goldstone mode associated with the breaking of the gauge symmetry. 
This simply reflects the fact that the charge is not a well-defined quantum number in a BCS condensate\cite{Senthil SQHE}. 
On the other hand, 
the residual U(1) degree of freedom left of the SU(2) spin rotational symmetry suggests that the transport of 
$S_{z}$ has universal feautures. This expectation is verified by gauging the 
system via the coupling to a spin-gauge field, from which one confirms  
that the 
spin Hall conductivity $\sigma^{s}_{xy}$ is quantised in integer multiples of $1/2\pi\hbar$. The integer factor is a topological invariant related to the 
chirality of the superconducting state, characterized in terms of the 
quantity ${\bf g}={}^{t}({\rm Re}d_{z}\tau_1 , -{\rm Im}d_{z}\tau_1 , \epsilon{\bf 1})$  as 
\begin{equation}
N_{chiral} =\int \frac{d^{2} {\vec k}}{(2\pi)^2} 
\frac{
{\rm tr}\left(
{\bf g}\cdot\nabla_{k_x}{\bf g}\times\nabla_{k_y}{\bf g}
\right)
%-{\rm g}_3 (\nabla_{k_x}{\bf g}\times\nabla_{k_y}{\bf g})_{3}
}
{
{\rm tr}\vert {\bf g}\vert^{3\over2}
},
\end{equation}
which is +1 for the present choice of d-vector (-1 for the opposite chirality).  To complete the discussion on symmetries and exotic topological properties,  
we note that vortices obeying nonabelian statistics can be realized in this chiral p-wave state
\cite{Ivanov}. 

%due to the 2-dimensionality, 
%a novel spin-disordered 
%chiral p-wave state as discussed by Zhou can be realized  with the 
%replacement ${\vec d}\rightarrow{\vec d}=(k_x +ik_y){\vec n}(\vec r)$ where 
%the spatial profile of ${\vec n}$ can result in a spin-texture supporting 
%zero modes of the quasiparticles. 

\noindent
{\it Relation to experimental inputs}

We list the salient features of the inputs provided by experiments 
conducted so far\cite{Sakurai}. 
\begin{enumerate}
\item{The normal state resistivity as a function of temperature 
fits rather well to a $T^2$-dependence. }
\item{Preliminary NMR measurement of $T_1^{-1}$ at the cobalt site 
shows a prominent peak just below T$_c$, reminiscent of the Hebel-Schlichter 
peak for conventional s-wave superconductors.}
\item{Spins direct their moment in-plane under an applied  magnetic field $\sim$ 8T.}
\end{enumerate}
Deferring discussions on feature 3 for the moment,  we focus on the first two. 
We take feature 1 to be an indication that the system lies in the 
weak coupling regime\cite{Rice}. It then becomes plausible to interpret feature 2
in terms of a fully-gapped pairing state without nodes. Although this is usually considered a hallmark of 
s-wave pairing, it is worth 
recalling that  
the Balian-Werthamer state,  
with ${\vec d}_{BW}=\Delta_{0}{\vec k}$ has a sufficiently strong singularity of the density of states to exhibit a 
coherence peak as well. The chiral p-wave state (with the cylindrical Fermi surface) 
discussed above is the 
two-dimensional analogue of this situation. An in-plane measurement of the phase-sensitive thermal conductivity 
on single crystals, when they become available, is expected to give an isotropic result.  
(The absence of a peak in the case of Sr$_{2}$RuO$_{4}$, which 
had also been considered 
to be a chiral-p wave superconductor, is consistent with the existence of 
horizontal nodes related to interlayer pairing\cite{Izawa}.) 
States such as 
${\vec d}=p_x {\hat z}$ and ${\vec d}={p_y}{\hat z}$, permissable on symmetry grounds,  can be 
dismissed on bases of these observations. 
Alternative fully-gapped states  
include the singlet $d_{x^2 -y^2}+id_{xy}$ and  
$d_{x^2 -y^2}+is$ states\cite{Baskaran}.     
One can directly distinguish between singlet and triplet states 
by performing a Knight-shift measurement. For singlet pairings, 
the shift $K$ will 
drop to zero on approaching $T=0$ for all magnetic field 
directions, with $K\propto Y({\vec k},T)\approx e^{-\frac{\Delta_0}{T}}$, where 
$Y({\vec k}, T)$ is the Yoshida function. Triplets on the other hand will  
exhibit an anisotropic magnetic susceptibility in the superconducting state;  
an in-plane field ${\vec H}\perp{\hat z}$ 
will polarize both the Cooper pairs and the 
quasiparticles, so that $K_{//}(T)={\rm const.}$ is expected, whereas the 
shift for ${\vec H}//{\vec z}$ will lead to a behavior similar to the 
singlet case, $K_{\perp}\propto Y({\vec k}, T)$. Remarkably, preliminary Knight shift measurements 
do show a temperature-independent behavior\cite{Sakurai}. (A large $H_{c2}\sim 61$T\cite{Sakurai} also is suggestive of triplet pairing.) 
These results, taken at face value single out 
the nodeless triplet pairing, of which the chiral-p state is the simplest.  

\noindent
{\it Possible scenario for triplet pairing in 
Na${}_{0.35}$CoO$_{2}$ $\cdot y$ H$_{2}$O}

According to LDA calculations\cite{Singh}, 
the central cylindrical Fermi surface of NaCo$_{2}$O$_{4}$, with dominant 
$a_{1g}$ character bears a shape close to a hexagon when 
viewed in the $k_z ={\rm const.}$ plane. 
Hence there are three (approximate) nesting vectors, given by 
${\bf Q}_i =\frac{4}{5}\Gamma K_{i}, i=1,2,3$. Here we define the
 $K$-points as $K_{i}=C_6^{i-1}(\frac{1}{3} {\bf G}_{1}+\frac{1}{3} 
 {\bf G}_{2})$, 
 ${\bf G}_{1}=2\pi\frac{{\bf b}\times{\bf c}}
 {{\bf a}\cdot{\bf b}\times{\bf c}}$, 
  ${\bf G}_{2}=2\pi\frac{{\bf c}\times{\bf a}}
 {{\bf a}\cdot{\bf b}\times{\bf c}}$, in which 
 ${\bf a}=(1, 0, 0), {\bf b}=(-1/2, {\sqrt{3}}/2, 0), {\bf c}=(0, 0, 1)$ 
 are the 
primitive vectors of the layered TL. $C_6$ denotes the 
anticlockwise rotation about the z-axis by an angle of ${\pi}/6$. The 
presence of such nesting instabilities in the $\Gamma -K$ direction has been suggested by  
Terasaki et al\cite{Terasaki SDW} 
based on their observation of an order-from-disorder type emergence of 
a spin density wave (SDW) in Cu-substituted 
NaCo$_{1.8}$Cu$_{0.2}$O$_{4}$. 
We are assuming below that these qualitative Fermi surface features 
remain intact for Na${}_{0.35}$CoO$_{2}$ $\cdot y$ H$_{2}$O, 
which appears to be in accord  
with the observed Curie-Weiss contribution to the susceptibility. 

Fermiology,  the study of many-body effects with an emphasis on Fermi surface properties 
has been playing a central role in advancing our knowledge on superconductivity in various 
correlated electron systems\cite{Aoki}.  
One such scenario\cite{Kuroki-Ogata-Arita-Aoki}, 
which takes advantage of nesting tendencies and leading to triplet pairing, 
has been advanced for the tetragonal lattice case, with the aim of explaining the pairing in 
Sr$_{2}$RuO$_{4}$. We would like to point out that a very natural extension 
to the present 
TL case is possible. This argument starts by 
recasting of the linearized BCS gap equation at T$_{c}$  into the 
form, 
$T_{c}\propto e^{-\frac{1}{N(O)<<V_{\phi}>>_{FS}}}$, where we are following 
the notations of ref.\cite{Kuwabara-Ogata}: 
$<<V_{\phi}>>_{FS}$ 
is the pairing interaction %term in the Hamiltonian 
$V({\vec k}-{\vec k}')$ %\phi({\vec k})\phi({\vec k}')$ 
averaged over the Fermi surface, i.e. 
\begin{equation}
<<V_{\phi}>>_{FS}=-\frac{
\int_{FS}d{\vec k}\int_{FS}d{\vec k}'
V_{\phi}({\vec k}-{\vec k}')\phi({\vec k})\phi({\vec k}')
}
{
[\int_{FS}d{\vec k}']\int_{FS}d{\vec k}\phi^{2}({\vec k})
},
\end{equation}
with $\phi({\vec k})$ the 
${\vec k}$-dependent part of the order parameter, i.e. 
$\Delta({\vec k})=\Delta_0 \phi(\vec k)$. 
We may view this as 
a sort of 
variational principle in which the actual pairing occurs for cases 
which yield a (1)positive and (2)large $<<V_{\phi}>>_{FS}$. In the present situation, where we are dealing with AF fluctuations, the pairing potential 
$V({\bf Q}_{i})>0$. For this to lead to pairing requires that 
$\phi({\vec k})\phi({\vec k}+{\bf Q}_i)<0$ for some $i$. The simplest way to realize  
this is to introduce a nodal direction 
which coincides e.g. with the $k_x$ axis, i.e. 
a $p_y$-pairing state. One immediately sees that this 
choice is highly beneficial as all three 
nesting vectors satisfy the required condition (see Fig. 1(a)), and therefore the entire 
Fermi surface is available for the pairing. (For this $p_y$-pairing, the 
${\bf Q}_1$-channel should be the most dominant because it coincides with the direction in which the 
magnitude of the gap becomes maximal.) In this respect, the advantage of such node formation 
(vertical nodes on the central Fermi surface) is bigger than in the ruthenates where only one pair of nested segments 
is involved in the pairing. 
This state is degenerate (i.e. has the same T$_c$) with 
those in which the nodes are located at the intersections between the 
hexagon Fermi surface and the lines $k_{y}={\pm}{\sqrt{3}}k_{x}$ (Fig. 1(b) and (c)). 
Further gain in condensation energy is achieved by constructing a linear 
combination of these three states in such a way that the nodes will cancel 
one other, opening a full gap at all points on the Fermi surface. 
This becomes possible through  
the construction 
\begin{equation}
\phi({\vec k})_{p_y +ip_x}
\propto\phi_{p_y}(\vec k)+i\left(\phi_{p_y}(C_{6}{\vec k})
+\phi_{p_y}(C_{6}^{2}{\vec k})\right).
\end{equation}
\vspace*{-8mm}
\begin{figure}[h]
\epsfxsize 8.3cm
\centerline {\epsfbox{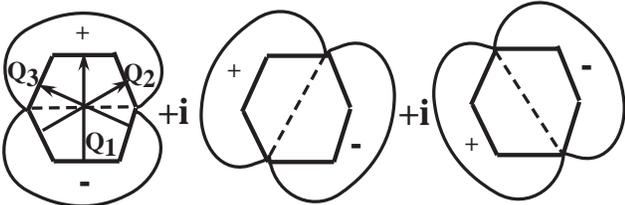}}
\caption{Illustration of pairing via nesting of Fermi surface segments. 
(a) Node located at $p_y$=0, corresponding to $p_y$ pairing. (b) and (c) denote the case where the nodal directions 
are rotated by 60 and 120 degrees respectively. Superposition of the three states, with relative phases of 
$\frac{\pi}{2}$ between (a) and (b), (c) will cancel the nodes, leading to a chiral p-wave state with enhanced 
condensation energy.  }
\end{figure}
\vspace*{-3mm}
As will be discussed elsewhere, one can check that a similar construction cannot be made for an f-wave state.  
The appearance of the chiral p-wave state above relied on the 
Fermi surface geometry (reflecting the hexagonal symmetry), and the 
assumption of a nesting tendency. It is desirable that 
fine quality samples, amenable to detailed neutron scattering measurements 
be made in order to make a direct verification on
this point possible.  Further support for 
triplet pairing comes from the aformentioned magnetic anisotropy;  when we interpert this in terms of an  
easy-axis along the c-axis,  evaluation along the lines of 
ref.\cite{Kuwabara-Ogata} can be used to demonstrate that the anisotropy ratio of the 
in-plane and longitudinal susceptibilities will directly enter into 
expressions for the triplet and singlet pairing potentials, and a sufficient 
anisotropy will eventually favor triplets (with ${\vec d}//{\hat z}$) over singlets.

We are indebted to H. Sakurai, K. Takada, T. Sasaki, M. Arai 
and E. Takayama-Muromachi for access to their results prior to publication, and for 
extensive discussions.  We would also like to acknowledge K. Kindo, K. Yoshimura and their group for generously 
allowing us to mention their results on magnetization and NMR. 
The similarity of the present system 
with the ruthenates was pointed out to the authors by I. Terasaki. 
We have also meritted from discussions with  T. Hikihara, T. Miyazaki, T. Sasaki, Y. Tateyama, K. Izawa, Y. Matsuda and J. Goryo. 

%\begin{thebibliography}{References:}

%\end{figure}
\end{multicols}
\end{document}